\documentclass[apjl]{emulateapj}

\usepackage{graphicx}
\usepackage{subfigure}
\usepackage{natbib}
\usepackage{amsmath}
\usepackage{bm}

\def \fermi{FERMI }

\def \be {\begin{equation}}
\def \en {\end{equation}}

\def \mtt {}

\def \ff{  }

\def \aq { }
\def \ac { }
\def \az { }

\def \aa { }

\def \nnn {}
\def \nnb {}
\def \rr  {}

\def \qq {}

\def \c {3C~279 }

\def \pg {PG~1553+113 }
\def \pgp {PG~1553+113}

\def \fermilat {{\it Fermi}-LAT }

\slugcomment{Accepted by the \textit{Astrophys. Journal} on Dec.
21, 2017.}

\begin{document}



\title{\bf The Blazar PG 1553+113 as a Binary System  of Supermassive  Black Holes}

\vspace{2.cm}
\author{
  M.~Tavani\altaffilmark{1,2,3,4}, A.~Cavaliere\altaffilmark{1,2},
  Pere Munar-Adrover\altaffilmark{1,5}, A. Argan\altaffilmark{1}}

\altaffiltext{1}{INAF/IAPS--Roma, via del Fosso del Cavaliere
 100, I-00133 Roma, Italy}
\altaffiltext{2}{Astronomia, Accademia Nazionale dei Lincei, via
della Lungara 10, I-00165 Roma, Italy}
 \altaffiltext{3}{Universit\'a ``Tor Vergata'',  Dipartimento di
 Fisica,
 via della Ricerca Scientifica 1, I-00133 Roma,  Italy}
 \altaffiltext{4}{Gran Sasso Science Institute,  viale Francesco
 Crispi 7, I-67100 L'Aquila, Italy}
\altaffiltext{5}{UAB, Barcelona, Spain}


 \vskip .1in



\begin{abstract}

The BL~Lac PG 1553+113  has been continuously monitored in gamma
rays  with Fermi-LAT for over 9 years. Its updated light curve now
includes  5  {\nnn iterations} of a main pattern comprising a high
peak and a longer trough, with a period P $\simeq2.2$ yr. Our
analysis of 2015-2017 data confirms the occurrence in January 2017
of a new peak fitting in the previous trend. In addition,
 we identify secondary peaks  (``twin peaks") that  occur in closely
symmetric pairs on both sides of {\nnn most}  main peaks,
including the last one; their  occurrence is supported by
correlated  X-ray outbursts.  We stress that the above features
strongly point to binary dynamics in a system of two black holes
(BHs) of some $10^8$ and $10^7 \, M_{\odot}$. At periastron the
{\nnn smaller} BH periodically stresses the jet $j_1$ launched by
the heavier companion, and triggers MHD-kinetic {\nnn tearing}
instabilities. These lead to magnetic reconnections and to
acceleration of electrons that produce synchrotron emission from
the optical to {\nnn X-ray} bands, and inverse Compton scattering
{\nnn into} the GeV range. For the origin of the twin peaks we
discuss two possibilities: a single-jet model, based on added
instabilities induced in $j_1$ by the smaller companion BH on its
inner orbital arc; and a two-jet model with the {\nnn smaller} BH
supporting its own, precessing jet $j_2$ that contributes lower,
specific {\nnn GeV} emissions. Such behaviors combining time
stability with amplitude variations betray plasma instabilities
driven in either jet by binary dynamics, and can provide a double
signature of the long-sought supermassive BH binaries.

\end{abstract}



\newpage

\section{Introduction}

{\aa The blazar \pg is a BL Lac {\az Object} at a redshift $z
\simeq 0.5$, detected in the optical band and at X-ray, GeV and
TeV energies.
%
{\nnn The first detailed study in the GeV range has been performed
by \citealt{abdo2010d}.}

{\aa {\nnn The} source recently attracted considerable attention
because of its clearly repetitive  gamma-ray emissions followed up
by \fermilat at GeV energies (\citealt{ackermann2015}, hereafter
A15). In fact, \fermilat was able to continuously monitor \pg over
a long stretch of time, from 2008 August 4 to 2015 July 19. This
6.9-year long data stretch reported in A15 showed a quasi-periodic
trend, with main peaks of gamma-ray emission occurring over a
period $P \simeq 2.18$ yr (observer frame, {\nnn corresponding to
$P/(1+z) \simeq 1.5$ yr in the source frame})
 at a confidence level greater {\nnn than $99$ \%}.
 Optical monitoring of the source shows prolonged
enhancements and emission minima in qualitative agreement with the
gamma-ray overall trend (A15).
 Furthermore, sparse X-ray monitoring of the source indicates the
 existence of  a more sporadic behavior of X-ray flaring not always correlated with
 the GeV main peaks.}

{\aa The 
 periodic GeV emission
admits an
interpretation in terms of {\nnn a} binary super-massive black
hole
 (SMBH)  system. {\nnn Establishing with certainty  the binary
 nature of a SMBH system such as \pg would have very significant
 implications concerning  study of
 blazar jets and BH evolution as advocated by e.g., \citealt{rieger2007},
 and would indicate  opportunities for future gravitational wave detections {\nnb from} SMBHs
 {\nnb by} projects such as the evolving LISA (cfr. \citealt{amaro-seoane2017}).}

 {\aa In the present paper we take up the challenge, and report
first our analysis of new publicly available gamma-ray and X-ray
data starting from July, 2015 until September, 2017. This
additional 2-year interval contributes considerable information on
\pgp. We therefore extend the analysis of A15 and confirm the
periodic nature of the GeV emission by determining the existence
of a $5^{th}$ main peak in the gamma-ray light curve that {\nnn we find} 
in January 2017. The timing of this peak is in agreement with the
epoch expected by A15 based on the period $ P \simeq 2.18$ yr.
Furthermore, we identify the existence of secondary  ``twin
peaks'' of  GeV and X-ray emissions that occur soon  before and
after the main GeV peaks. We present in Sect. 2 the results of our
analyses of the updated  data that support these findings.}


{\aa We then turn to  theoretical interpretations of these
emission
features}.  
 {\aa First we recall our context. BL Lac
Objects \ {such as PG 1553+113} constitute a subclass of  blazars.
These are marked among the quasars by narrow, relativistic
jets of electron-proton plasma {\nnn and} embedded magnetic field
{\nnn with a substantial axial component}. The jets are launched
by a central super-massive black hole (SMBH) with mass $M \, \sim
10^8 - 10^9 \, M_{\bigodot}$, \ {and flow upward of its accretion
disk} with
 bulk Lorentz  factors $\Gamma \sim 10 $.  Thus {\rr their radiations  are
 boosted and collimated} and  appear  so  bright as to "blaze"
the observer  when   aligned with the line of sight \ {within
 angles $\theta  \simeq 1/ \Gamma$.} \ {{\nnn On the other hand, many
 similar {\rr if inconspicuous} objects  (a number $\Gamma^2$ times larger)} are  expected
  {\nnn to exist} outside this angular range.}

The  observed outputs are \  {variable and } up to some $10^{47}$
erg/s in non-thermal radiations produced by highly relativistic
electrons with random Lorentz factors up to $\gamma_p \sim 10^3$
accelerated in the jet. In fact, BL Lacs \ {lack - or hide in the
{\nnn central throat of their thick accretion disk} - } thermal
features such as the broad optical lines and the Big Blue Bump
shining in many quasars including the other blazar subclass, the
Flat Spectrum Radio Sources.

 In BL Lacs the    spectral   energy distribution  (SED)  is
constituted  by two {\rr nearly level} humps: one  --  peaking in
the $IR \, - \,UV$ and declining toward  the soft X rays -- is of
agreed  synchrotron (S) origin; the other rises from hard X rays
to  the GeV band  with occasional bursts up to  the TeVs, and is
of likely inverse Compton (IC)  nature  (for basics of radiative
processes  see \citealt{rybicki1979}).

Detailed SED modeling 
 ({\qq e.g.,
\citealt{cavaliere2017}, hereafter CTV17}, Table 1)
confirms that both humps are produced when relativistic electrons
with number
 densities $n \sim 10^3 $  cm$^{-3}$  interact with the jet magnetic fields
$B  \sim 1$ G to  emit  S  radiation at frequencies $\gamma^2  \,
eB/mc$, and produce from the source size $R \sim 5 \cdot 10^{16}$
cm observed  (isotropized)  luminosities  $L_S \propto n \, R^3 \,
B^2 \, \Gamma^4 $. Meanwhile, by  IC  in its S-Self Compton
version \citep{maraschi1992} suited to BL Lacs,  the relativistic
electrons upscatter the very S photons to energies $\gamma^2$
{\mtt times higher} (full discussion is given, e.g.,  by
\citealt{urry1995, peterson1997, ghisellini2016}).


\begin{figure*}
\vspace{1cm}
   \centerline{\includegraphics[width=17cm, angle = 0]{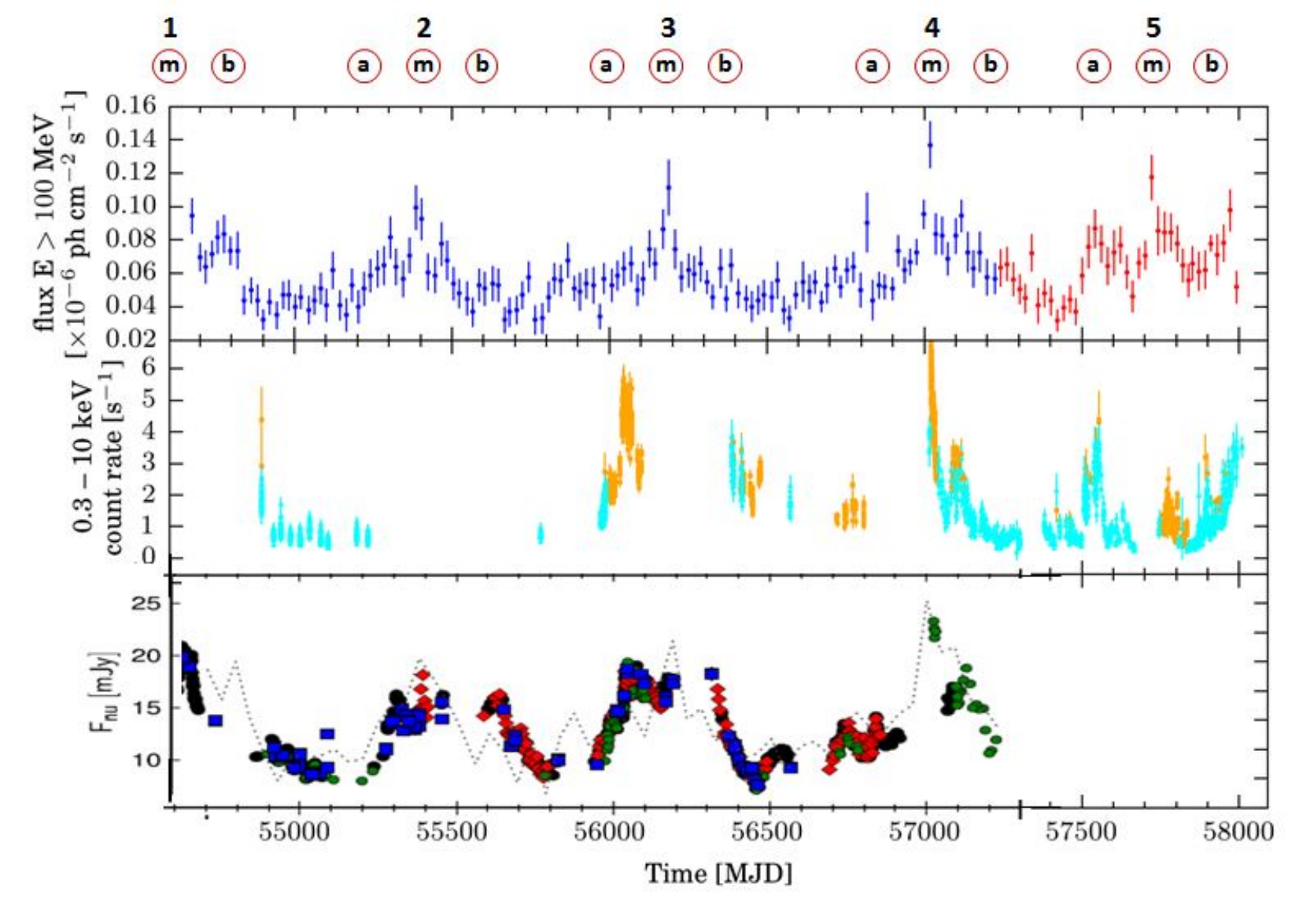}}
 \caption{Light curves of \pg observed throughout 2008 - 2017
in various bands. Top panel: gamma rays  from Fermi LAT, with the
2016 - 2017 data traced in red, main peaks marked by
\textcircled{m}, and twin peaks marked by \textcircled{a} and
\textcircled{b}. Middle panel: keV X rays from SWIFT. Bottom
panel: optical R band reported from Fig. 2 of
\citealt{ackermann2015}.}
\end{figure*}

\vspace{0.3 cm}
\section{Analysis of updated gamma-ray data}


The A15 analysis identified a  quasi-periodic behavior of the
gamma-ray emission of \pg based on 4 main peaks. Their analysis
spans the time window from July 2008 until July 2015. We extend
the analysis to include the time interval up to September 2017.
This additional 2-year stretch of data {\nnn (shown in Fig. 1)}
provide important information to confirm the gamma-ray periodicity
and add new clues.
%
{\mtt Details of the data analysis of \fermilat data are provided
in the Appendix A1.}

As anticipated  above, a new peak has occurred at the expected
time based on the 2.18 year periodicity of A1. The updated  set of
data now provides a full 5-fold {\nnn iteration} of a main pattern
comprising a prominent peak and a prolonged trough. We give in
Fig. 1 the complete gamma-ray light curve with the new data marked
in red color.

To analyze periodicity and additional features in the signal we
used  tools that do not refer to any sinusoidal pattern but rather
adopt data folding and shifting (phase dispersion minimization,
PDM, see Appendix A2) and projections onto a {\nnn localized}
vectorial basis such as the continuous wavelet transform (CWT, see
Appendix A3). Our results {\rr concerning \pg} are presented in
Fig. 2 and in Figs. 3 - 4,  respectively. {\nnn We have checked
the red noise properties of the gamma-ray data, and   found they
can be represented by a decreasing power-law with index  $0.83 \pm
0.11$ consistent  with A15.  The related confidence levels  are
reported on Fig. 4}. From our analysis we conclude {\mtt that} the
\textit{main} cyclic pattern is confirmed with a period $ P \simeq
2.2 \, \rm years$ {\nnn at a confidence level significantly
greater than  $99$\%}.

 {\nnn On the other hand, we also} detect  a
\textit{secondary} pattern characterized by pairs of lower peaks
(``twin peaks'') flanking the main ones as we discuss
in Sect. 4. Such features are 
{\mtt clearly} {\nnn pinned down by the CWT algorithm in most
cycles, and especially at around MJDs 54800, 57550, and 57950}
with increasing associated powers (see Fig. 3 and its caption).
 In fact, a  close look at
 the gamma-ray curve in Fig. 1 (top panel) confirms these features,
including their increasing amplitudes. Our {\nnn Fig. 4 shows the
CWT  power obtained for two specific temporal sections. A first
one during the cycle 3 without strong twin peak signal (black
curve), and a second one during cycle 5 with a twin peak signal at
confidence levels greater than $95$\% (red curve).}

We mark in Fig. 1 by "a" and "b" the times of actual or expected
occurrences of twin peaks, and  discuss the
phenomenon in Sect. 4. 
Such an updated and rich data set warrants a new and aimed
discussion.







\vspace{0.3 cm}
\section{Interpreting the main periodic pattern}

{\mtt We base our interpretations on } the dynamics of a binary
supermassive black hole (SMBH) system comprised of a mass $M_{1}
\sim 5\ 10^{8} M_{\bigodot}$ and a companion of {\mtt smaller}
mass $M_{2} \sim M_{1} /10$,   tracing closely Keplerian orbits
around the center of mass at an average distance of a some mpcs
(see A15).

A  \textit{geometrical} interpretation is based on periodic
visibility within the $ \theta \sim 5^{\circ}$  cone of jet
emission, as detailed by \citealt{sobacchi2017}. However, such a
model not only is  limited to the S radiation components, but also
is  constrained to describe the oscillations of a secondary jet
$j_{2 }$ launched by the {\mtt smaller mass} BH,
lest {fast dissipation of } orbital energy into gravitational
waves would cause early onset of orbital decay and catastrophic
merging (same authors, their Eq. 28). But then, {there is no
trace in the data} of the emission closely perpendicular to the
orbital plane from the jet $j_{1}$ associated with the major
component $M_{1}$; yet this ought to be dominant due to its higher
accretion rate and luminosity, typically  $L_{1 } \propto M_{1}$
in the Eddington regime.

Our interpretation  (taking up from  CTV17)  is instead aimed at
high-energy radiations and electrons, and is\{centered} on
\emph{dynamically} triggered  evolution of the
primary jet $j_{1}$ (see Fig. 5). We recall  that  in a blazar of
the BL Lac type like \pg such a jet with its {\az intermediate}
magnetization $\sigma _{j} = B^2/4 \, \pi \, n \, m\, c^2 \,
\Gamma \approx 1$
({\rr the ratio of magnetic and bulk kinetic energy densities in
the observer frame}) is \textit{metastable}. Large scale, MHD
instabilities are triggered by the varying gravitational force $F
\propto 1/r^{2}$ exerted by $M_{2}$ \ {from a distance ${r}$ (with
$r \simeq r_2$ to within $0 ( r_1/r_2 ) = M_2/M_1 \ll 1)$ along }
its \ {Keplerian } elliptical orbit with eccentricity $\epsilon
\simeq 0.1$; specifically {\mtt  on a critical orbital arc around
periastron that covers  about 20\% of the total (see Fig. 2 of
CTV17)}.

Thus the magnetic {\bf B} lines are compressed, distorted and
twisted, and so made prone to collisionless tearing and
reconnecting instabilities that proceed down to the kinetic level.
Then the induced \textbf{E} fields accelerate electrons to
energies that attain Lorentz factors $\gamma \sim 10^{3}$  for
values of the {\az local} electron magnetization $\sigma_{e} \
\propto  \ \sigma_{j} \ m_{p}/m_{e} \gg 1$. Such a generic kind of
evolution with its MHD - kinetic \textit{transition} has been
widely computed, discussed (see, e.g., \citealt{mignone2013,
kagan2015, striani2016, yuan2016}), and applied by
\citealt{petropoulou2016} to conditions expected to prevail in
blazar jets.

Focusing   on  \pgp,
{\nnn we argue } that in fields $B\sim 1$ G the high energy
electrons accelerated within the jet
 at heights around $10^{16}$ cm  will emit  S radiations from the
optical to the keV X-ray band (and occasionally beyond);
meanwhile, such photons are upscattered into the GeV range after
the SSC radiation process. This view  is in tune with the complex
correlations gamma-ray -- X-ray -- optical  present in the {\mtt
updated } data (compare Figs. 1: top, middle, and bottom panels),
once the following expected features are considered: optical and
X-ray synchrotron emissions occurring {\nnn in the axial}
\textbf{B} at increasing heights and photon energies $E \, \propto
\, \gamma^{2 }$ will be increasingly focused into angles of order
$1/E^{1/2 }$ at the single electron level that \ {constrains from
below the overall} beam widths.

In fact, the {\nnn optical}  emissions are observed to be quite
smoother and broader \ {when compared to} the \textit{spiky} X
rays; \ {the latter feature narrow high-energy surges (likely
powered by fluctuations in the energy distribution tail)} that may
be easily missed. Instead, gamma rays from IC scattering cover a
broad
emission fan that can be continuously observed, and
sometimes reach the VHE range (e.g., \citealt{aleksic2015}).

We stress that the basic peak/trough  pattern \ {repeating
itself throughout the gamma light curve} of \pg features a \
{sharp aspect ratio}, as expected
from MHD - kinetic
instabilities \emph{triggered} by the gravitational \emph{force}
$F(r) \propto 1/r^{2}$ near periastron \ {and then \emph{relaxed}
along the rest of the orbit.} \ {Whence we conclude that in the
main pattern moderately  variable \emph{amplitudes}
combine with precisely preserved \emph{timing } to constitute the
hallmark of jet plasma physics driven by binary dynamics as
occurring in \pgp.}

\vspace{0.3 cm}
\section{The secondary twin peaks}

%

We have anticipated in  Sect. 2 the  evidence we find in the
updated gamma-ray data concerning  twin secondary  peaks that
flank in a nearly symmetric disposition most main peaks of the
basic periodic pattern, and last for some  $ 10^2$  days.

Here we stress how extant such  substructures are  in cycle 5.
Similar features also loom out {\nnn with } a lower amplitude near
MJD 55000 at the very beginning of the \fermilat observations that
open cycle 1; then they come in full view in  cycle 2 and in cycle
4. In all such instances  their  presence and quantitative
parameters are strongly supported by our CWT analysis (see Fig. 3,
4 and their  captions) in the form of elongated spots between
about MJD 57000 and MJD 58000 with a {\mtt time interval of} $\sim
300$ days, that corresponds to the {\mtt occurrence of twin peaks}
in the light curve.


In conclusion, the twin peaks appear with similar parameters in
our  analysis tools: light curves and CWT. {\aq Thus they call for
an explanation. In the following, we discuss two possible ways
toward understanding their production: a single-jet interpretation
based on additional instabilities induced in the primary $j_1$;
and a two-jet interpretation including the contribution of the
independent relativistic jet $j_2$ launched by the secondary BH.}

\begin{figure*}
\vspace{1cm}
   \centerline{\includegraphics[width=12cm, angle = 0]{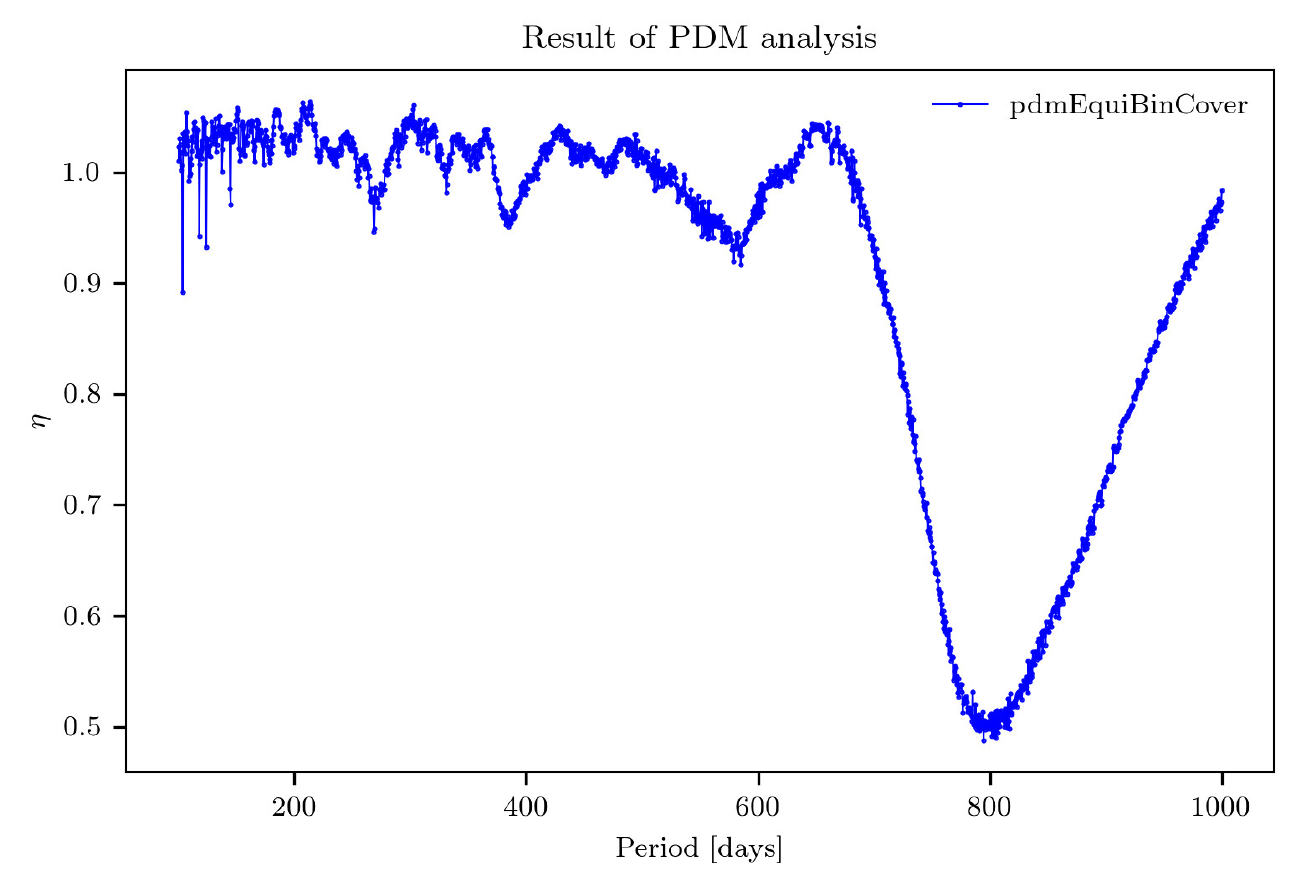}}
 \caption{PDM \ {(Phase Dispersion Minimization)} plot related to gamma-ray Fermi-LAT
  data of $\pgp$  from 2008 August 4 to 2017 September 4.
  The plot features a {\nnn deep}, sharp minimum at the period P = $2.18 \pm 0.03$ yr.}
\label{fig-2}
\end{figure*}


\begin{figure*}
\vspace{0.5cm}
   \centerline{\includegraphics[width=16cm, angle = 0]{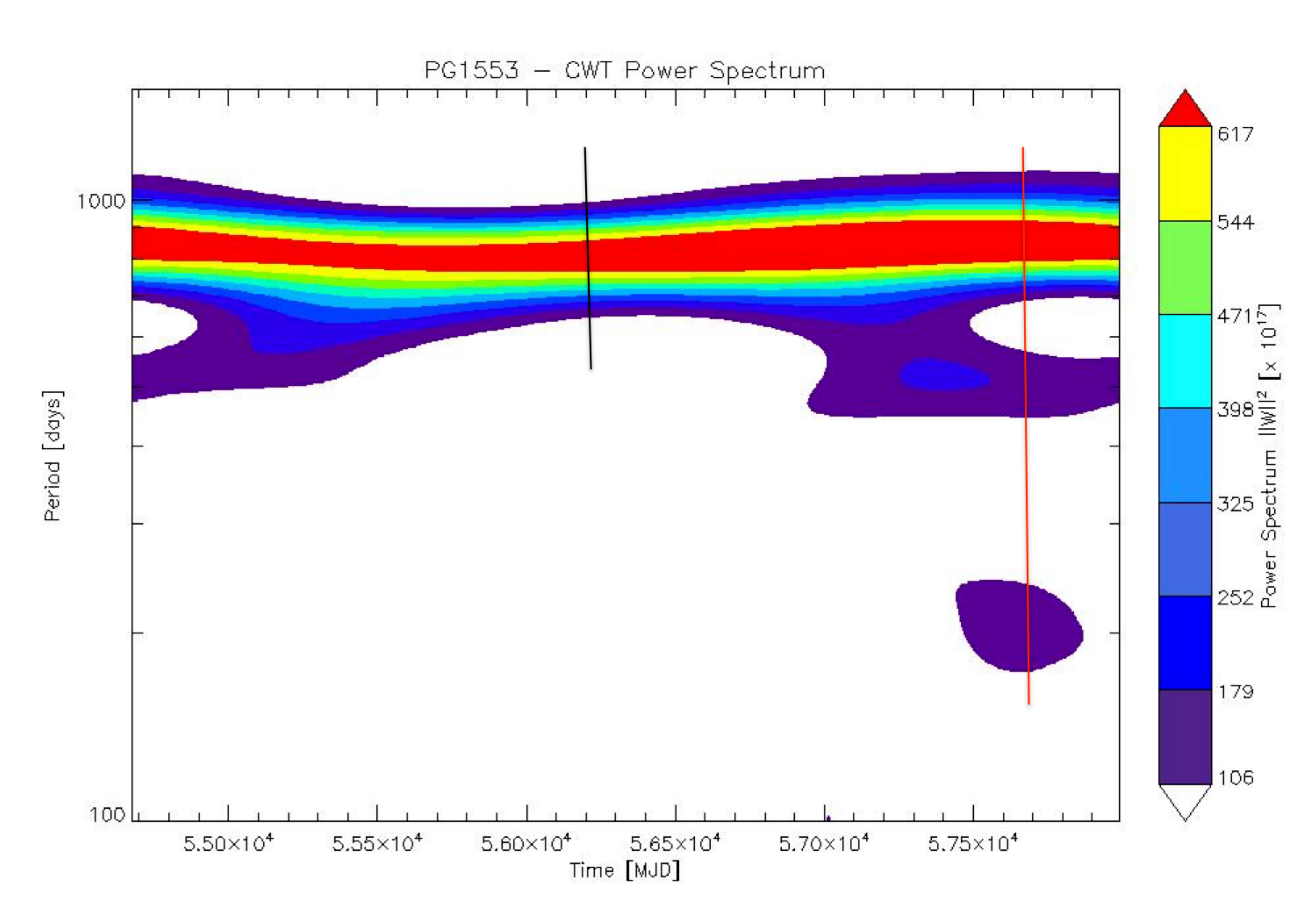}}
 \caption{ CWT (Continuous Wavelet Transform) synoptic map based on the standard
Morlet 6 wavelet, of period and {\nnn spectral} power versus time
of the Fermi-LAT data {\rr concerning \pg} from 2008 August 4 to
2017 September 4. The power spectrum shows an  extant stripe
at normalized amplitudes around 650  that supports the main
periodicity of $P \simeq 2.2$ yr.
{\nnn It also} shows spots at amplitudes around 140 and lower,
 at times around MJD $\simeq 55300,   \, \, 57200,  \, \, 57700$
related to secondary structures.
In fact, looking into the updated gamma-ray light curve of Fig. 1
(top panel), a closely repetitive secondary feature is recognized
in cycles  2, 4, and 5
in the form of twin peaks flanking the main one at times around $
\Delta t \simeq \pm \,250$ d. {\nnn The two vertical lines
correspond to the time sections used for the analysis reported in Fig.
\ref{fig-4}}.}

\vspace{1cm}
   \centerline{\includegraphics[width=12cm, angle = 0]{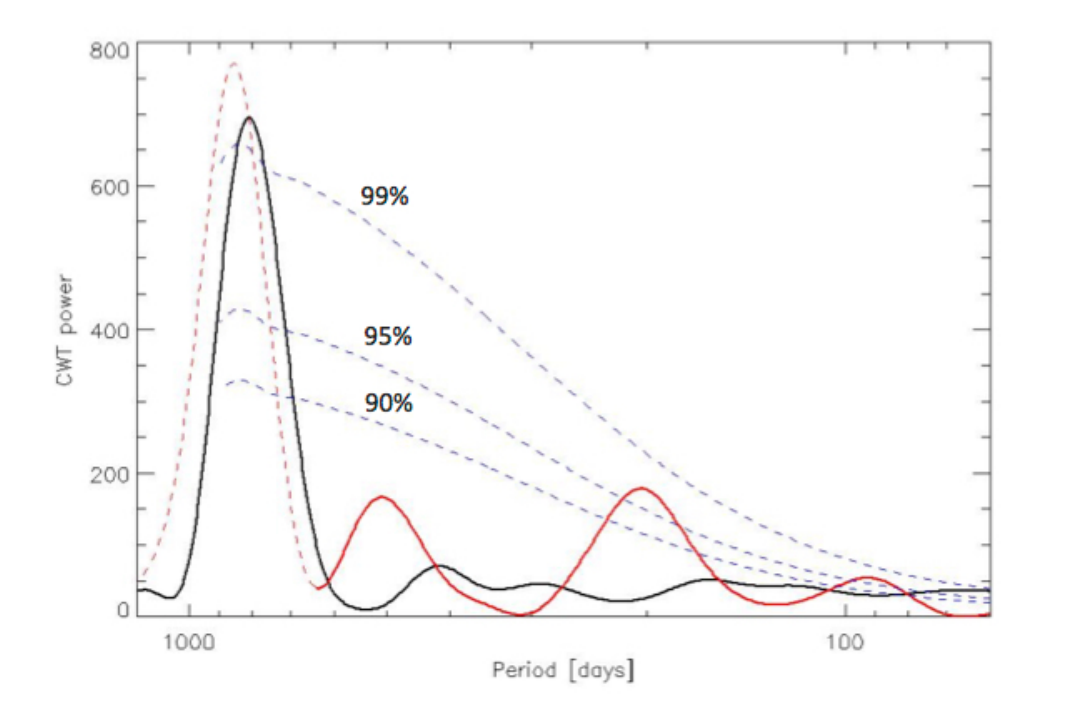}}
 \caption{{\nnn The CWT power of the gamma-ray lightcurve determined for
 two time sections (as {\rr highlighted} 
 in Fig. 3). The black curve shows the power for a time
 section in cycle 3 selected for the absence of twin peaks. The red curve provides the wavelet power for a time
 section in cycle 5 where  the twin peaks phenomenon is
 strongest. Dashed curves show {\rr specific} confidence levels obtained from red noise with power law
 index 0.83  (see Sect. 2).}}
\label{fig-4}
\end{figure*}

\vspace{0.3 cm}
\subsection{A single-jet interpretation}

This interpretation is based on a development of the discussion
in Sect. 3
concerning the {origin } of the main periodic pattern, to cover
also existence and features of
 the twin peaks in the light curve of \pgp. The basic driver is again identified in
substantial magneto-gravitational stresses induced by the orbiting
companion BH $M_2$ that induce strong  perturbations in the
structure of the main jet $j_1$ launched by $M_1$.


The process of producing the twin peak phenomenon has to explain
the {very presence} of twin peaks in the gamma-ray light curve,
and their approximate time symmetry relative to a main one, Other
points to note are the correlation with  X-ray spikes and the
often enhanced states of optical emission.

The repetitive, yet not fully periodic,  nature of the twin peaks
suggests an interpretation in terms of an \emph{unstable} source
of additional emission that is superimposed to the mechanism
underlying the main periodic pattern in
gamma rays (and related optical emissions) associated with the
main peaks. Here we emphasize that the gravitational perturbation
of $j_1$ induced by the companion BH as it transits the critical
orbital
arc 
can produce a \emph{number} of MHD transient structures. Fig 5
(left panel) provides an outline of the related configuration.

{\ff As the secondary BH of mass $M_2$ enters the critical arc,
magneto-gravitational stresses are induced in $j_1$ and propagate
longitudinally along the envelope of the magnetic helicoidal field
threading the jet.
{\mtt Helicoidal and shear MHD modes \citep{buratti2012}} can be induced by
this process. This  induces magnetic twisting and reconnections in
$j_1$ {\mtt as detailed in Sect. 3}.
The resulting
tail of the electron energy distribution is {enhanced} by the
presence of such an additional source of accelerations. }

{\ff In closer detail,  we estimate at $\kappa \, R_{S}
\simeq 10^{15}$  cm (in terms of the Schwarzschild radius $R_S
\simeq 10^{13} \, \rm cm$ and of the factor $\kappa \sim 10^2$)
the radius of the inner region of the accretion disk surrounding
$M_1$ wherefrom  the $j_1$ jet is
launched. The gravitational force 
induced by $M_2$ on
the plasma and magnetic structure of $j_1$ will
enhance MHD
shear 
 modes with low azimuthal index $m$ that
propagate along the jet axis. We note that the timescale for such
propagation  is bounded by $\Delta L /c$, when the Alfven velocity
is close to $c$. We therefore obtain a coherence {\mtt scale} shorter than
$10^2$ light-days corresponding to the width of the twin peaks for
$\Delta L \sim  10^{17}$  cm, a natural scale for BL Lac jets. }

{\ff Eventually, the additional MHD perturbation fades away, and
the plasma column of $j_1$ relaxes to a lower energy state as
typical of sub-critical plasma  configurations perturbed by
external drivers (for laboratory tests, see
\citealt{buratti2012}). In this interpretation, the secondary peak
preceding the main peak is induced by the very entrance of the
secondary BH in the critical orbital arc. The secondary peak
following the main one is produced at the very end of the process,
as the companion BH leaves the critical orbital arc and $j_1$
reverts back to its unperturbed state}.

The effect of the external, transient perturbations is reflected
in the X-ray range, where the S emission is dominated by top
energy electrons.  X-ray spikes observed in \pg
during cycles 4 and 5 are a likely product of the mechanism.
Meanwhile, inverse Compton radiation from the same electron
population produces the twin peaks in the gamma-ray range. We
speculate that the radiative outcome may extend into the {TeV}
range,  {as far as allowed by the Klein-Nishina cut off. } The
available optical light curve of \pg (see A15 and Fig. 1, bottom panel) shows
episodes of enhanced  emission that correlate with the general
trend of increased gamma rays, and features  highs (if not flares)
corresponding to twin peaks.

The potential importance of X-ray emissions in this context is
stressed by the detection of X-ray flares correlated  with the
gamma-ray twin peaks of cycle 5, specifically at MJD~$\sim$~57530
and 58000 (see Fig.~1, middle panel). In addition, despite the
sparse X-ray monitoring of the source during cycles 1-4, a very
prominent X-ray outburst near MJD 56050 was detected (with little
gamma-ray response) at a phase typical of a twin peak.

Natural consequences of the process include: an intrinsically
{\nnn unstable} behavior of the twin peak phenomenon in the
gamma-ray range, as caused by the {\mtt turbulent} nature of the
driving MHD instability; X-ray outbursts corresponding to maxima
of twin peaks of similar strength as for the main peaks (see cycle
5), and the secondary peak episode in cycle 4; lack of systematic
correlation between optical highs and  X-ray outbursts.

The last point  is important for BL Lacs. Indeed, lack of
correlation between the X-ray and optical emissions has been often
observed in {several such sources } such as Mrk 421  \citep{balokovic2016}.
In the present case, the periodic nature of  \pg
provides the first evidence for a prime driver of electron
acceleration whose recurrent pattern can be used to model the
properties of relativistic jets of BL Lac blazars.
%
%

\vspace{0.3 cm}
\subsection{Two-jet interpretation}

{\aq We now consider a two-jet interpretation based on emissions
also from the jet $j_2$  associated with the secondary BH (see
Fig. 5, right panel).} {\az An  origin in $j_{2}$ of the  twin
peaks  is indicated by their specific features relative to the
central main peak: the definitely lower amplitudes by a ratio
about $1/2$ or less; the lack of detailed amplitude correlation
with the main peak's; their own, shallow peak/trough pattern. }

The first  feature constitutes the positive, telling side of what we
noted in Sect.3;  namely, a \emph{weaker} {\ff accretion power}
is available for twins from the lower accretion rates
prevailing at the base of the secondary jet $j_{2}$. In fact, the power
will  scale as $L_{a}, \, L_{b} \propto M_{2} \ll M_{1}$.

{\ac The  second and third features relate in the present view to
the specific \emph{trigger} of the twins' emissions.}
{\ac They would be produced not directly by the sheer
gravitational force from $M_1$, but rather  by  the \textit{torque
T } constituted by the latter } coupled with jet base reaction.
 {\ac This is because  a torque's action is due
to be smoother, } since it is proportional to $T(r) \propto r \,
F(r) \propto  1/r$ .

  In detail, the vectorial  torque acting on
the mass $m_{2}$ of the secondary jet is given by $\boldsymbol{T}  = \boldsymbol{F}
\wedge \boldsymbol{\ell}$ in terms of the gravity force $F = G \,
m_{2}\,  M_1 /r^2$ with its lever arm $\boldsymbol{\ell}$,
induced by the massive companion and dominating the local gravity
proportional to $  M_2 /\ell^2$. The modulus $F \, \ell \, \sin
\Theta$ includes an angular factor that may be approximated as
$\sin \Theta \simeq r/\ell\, $ for $\, r/\ell  < 1$. This implies
a \textit{softer} change of dynamical stress along the orbit
caused by the torque $T (r) \propto 1/r$, resulting in a
\textit{smoother} peak/trough secondary pattern than
could arise from the force itself.

\ {On the other hand, the maxima of the twin peaks $a$ and $b$
occur at a time distance $\Delta t_{a}$ and $\Delta t_{b}$  from
their central main peak at $t_i$,  {\nnn and outside its
half-width} $\delta t_i $. The observed nearly \textit{symmetric}
locations $\Delta t_{a} \simeq \Delta t_b >\sim
\delta t_i $ (as indicated on Fig. 1, top panel) call for a torque
action with maxima not really simultaneous to the maximal force,
yet inducing $j_2$ visibility close to the beginning and the end
of the strong-force range around periastron, see Sect. 4.

Guided by such a detailed  observational evidence,  we submit that
the torque specifically causes $j_{2}$ to bend/release into/out
the visibility cone, while also triggering the jet's internal
instabilities  {as } \ {discussed in Sect. 2}. Note that our
interpretation implies that a twin member can appear close to a
main peak \emph{both} on its ascending and descending shoulder, as
in fact observed.


{\nnn Next,} we discuss key parameters and quantitative relations
{\nnn in the light of }  a simple model for $j_2$; here a toy-top of
substantial angular momentum and high  angular velocity $\omega$
(limited only by $ c/2 \pi R_{S2} \simeq 5\, 10^{-3} $ s$^{-1}$,
in terms of the Schwarzschild radius $R_{S2} \sim 10^{12}$ cm
$\propto M_2$) undergoes a slow precessional rotation with
velocity $\omega_{p} \ll \omega$ under the action of a torque (see
Fig. 5-5 in \citealt{goldstein1959}).
%
 When averaged over nutations,
\ {the dynamics of rigid bodies yields the  precessional velocity
in the simple form (cfr. \citealt{goldstein1959}) }
\begin{equation}
\omega_{p}  =  T / \omega \,  I_{2}  \, , \, \,
\end{equation}
\noindent in terms of the torque $T$ divided by the  angular
momentum $\omega \, I_{2}$, given the  moment of inertia $I_{2}
\propto  m_{2} \, R_{S2}^{2}$; in fact,  $m_{2}$  cancels out
between $I_2$ and $T$.


\ {In our binary system,  an orbital phase-depending torque
$\boldsymbol{T}$ is provided by the gravitational action from
$M_1$ coupled with the opposite reaction from the $j_2$ base. The
latter is likely constituted  (similarly to the $j_1$ structure
discussed in Sect. 4.1 above) by a central disk region with radius
$\kappa \, R_{S_2} $ larger than $R_{S_2}$ by a factor $\kappa
\sim 10^2$.
Then $I_2$ scales up by the factor $\kappa^4 \sim 10^8$; this
helps to achieve a substantial angular momentum in a standard
``proton-loaded" jet with total density {\nnn of some} $10^5$
cm$^{-3}$ and  radiative efficiency {\nnn around} $5$\% (cf.
\citealt{celotti2008}).

\ {Our main point is as follows}. If just  a few precessional
rotations occur (as if close to a resonance) over the strong
interaction arc that takes  some 10\% of the orbital period around
the periastron, we expect $\omega_{ p} \approx 10 \, \,
\omega_{0}$ (still $\ll \omega$) to hold for the precessional in
terms of the orbital frequency. Again  barring nutations, this
translates into the approximate values
\begin{equation}
\Delta t_{a} \approx  \Delta t_{b} \simeq \,  P/10\,
\simeq \, 3  \,\,\rm months .
\end{equation}
\noindent in agreement with the observed spacing of the twin peaks
in the light curve. The precession provides a dynamical memory
that enforces twin peaks symmetry {\nnn and also causes smooth
amplitude changes from cycle to cycle}.

We add that closer insights into $I_{2} $ (that is, structural
information concerning jet and disk of the minor companion) may be
obtained from the bounds
 on $\omega $ that scale with $M_{1}$ and $M_{2} $ to read}
\begin{equation}
\omega <  c/2\pi R_{S2} \propto  1/M_{2}, \, \, \, \, \, \, \, \, \, \,\omega
>\omega_{min}  \propto   M_{1}^{1/3} /M_{2}\,   P^{1/3} \, .
\end{equation}
\noindent Here we have used Eq. 1,  $T\propto M_{1}/r$, $I_{2}
\propto \, R_{S2}^{2}$ as said above, and the $3^{rd}$ Kepler's
law in the form $r\propto M_{1 }^{1/3} P^{2/3}$.

Clearly, deviations from rigidity are expected, and might cause
wide  variance of the above values. Actually, the data in Fig. 1
(top panel) show the twin peaks to have undergone just {\nnn a}
slow amplitude  increase  over the last few periods, yet
preserving  good \emph{permanence} of their relative timing
$\Delta t_{a} \approx  \Delta t_{b}$. Such a behavior apparently
constitutes yet another instance of the general trend: timing
permanence vs. limited
amplitude variations, that  marks jet plasmas affected by overall
binary dynamics  as noted at the end of {\nnn Sect. 3}.}

\begin{figure*}     [!htb]
\includegraphics[width=8cm, angle=0]{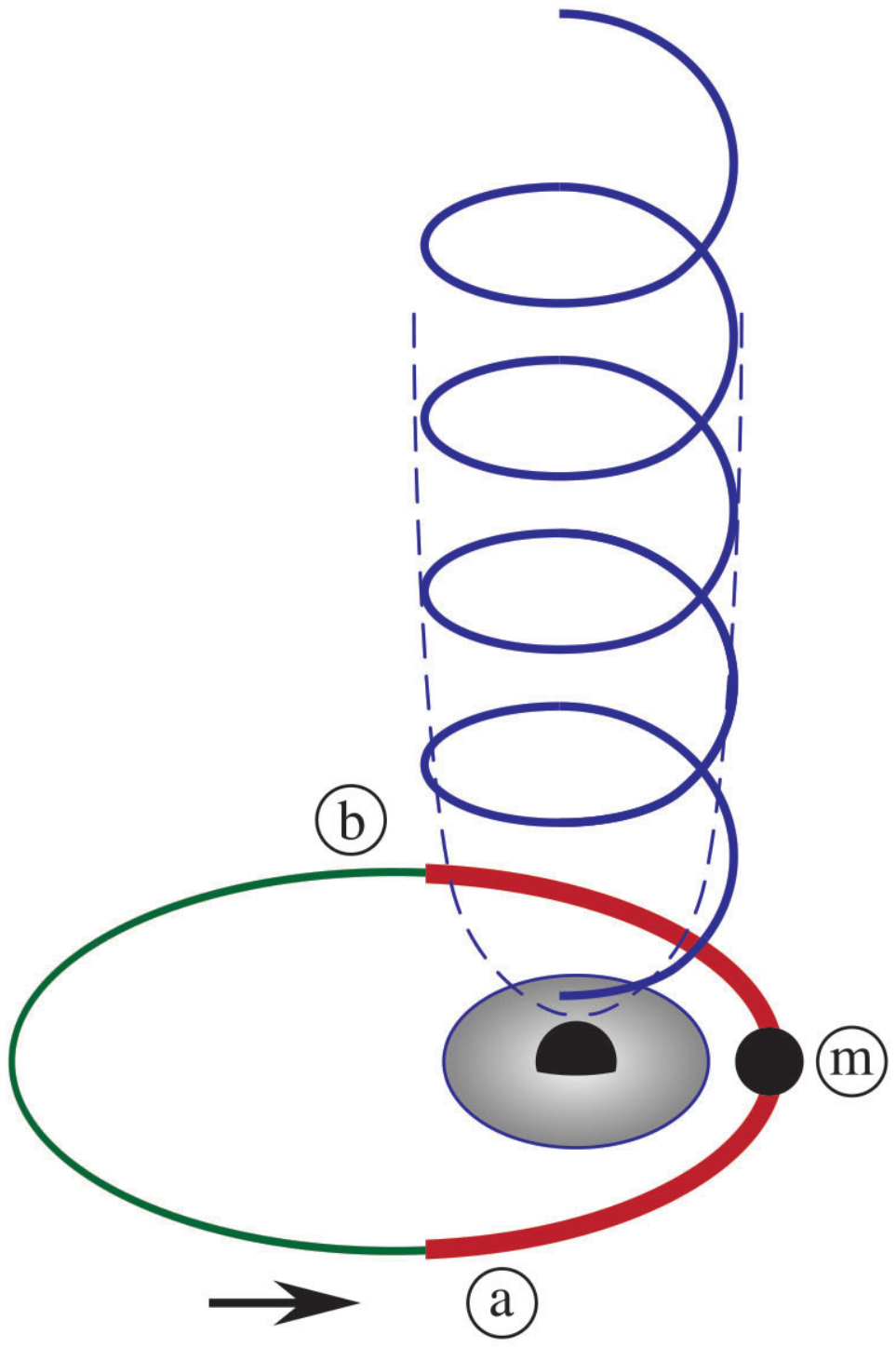}
\includegraphics[width=8cm, angle=0]{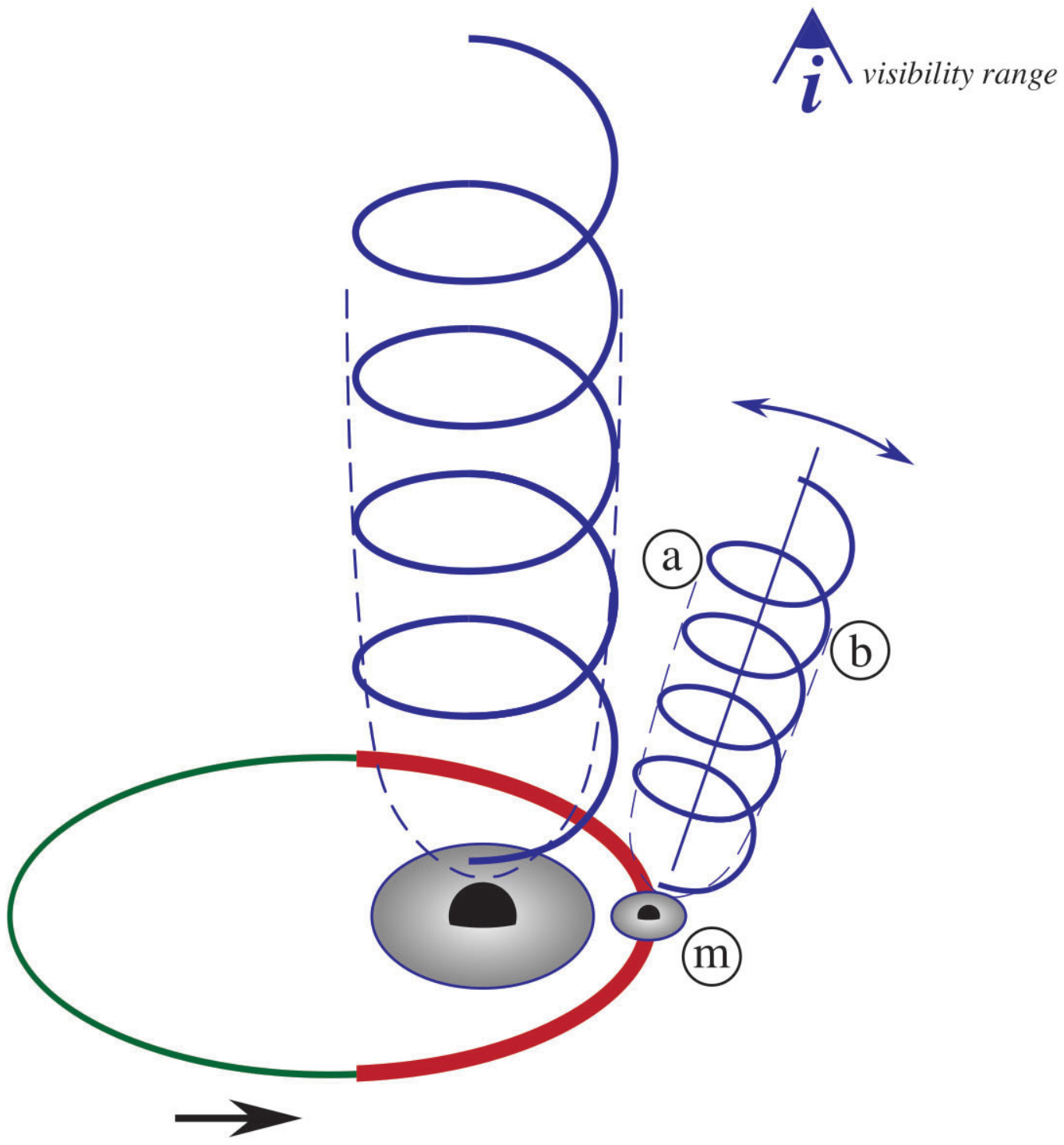}
\vspace{-3cm}
\caption{Left panel: schematic view of the SMBH binary \pg
according to
 the one-jet interpretation in Sect. 4.1.
 Right panel: view of the SMBH according to the two-jet interpretation in Sect. 4.2.
 The symbol \textcircled{m} marks the orbital position wherefrom the main peak is emitted, whereas the symbols \textcircled{a}, \textcircled{b} mark the positions wherefrom the twin peaks are emitted.
 The jets outflow with velocities $v = c \, (1- 1 / \Gamma^{2})^{1/2}$. }
\label{fig-5}
\end{figure*}

\vspace{0.3 cm}
\section{Conclusions}

The data concerning the emissions of PG 1553+113 as updated 
{\nnb toward} present  confirm the periodic behavior looming out
in A15, and add considerable new information. The updated light
curves warrant close analyses as we  carried out in Sect. 2,  and
detailed interpretations as we discussed in the rest of the
present paper.

Sect. 3 took up from CTV17 to address the origin of the main,
periodic peaks now observed throughout full 5 cycles in the
updated light curve of \pg ($z \simeq 0.5$), and {\nnb to} trace
them back to the \textit{dynamics} of a binary SMBH system.
Specifically, the main jet $ j_1$  is periodically stressed by the
gravitational force from the smaller mass BH, as the latter covers
a critical orbital arc (some 20\% of the orbit) across the
periastron.

The stress destabilizes the sensitive BL Lac-type main jet $j_1$,
by itself in a \emph{metastable} condition due to its relatively
large magnetization. 
The instability starts on large, MHD scales, then progresses into the
collisionless kinetic regime and accelerates electrons  to
relativistic energies. These power non-thermal radiations, namely,
synchrotron  in the O, X bands and inverse Compton  in gamma rays.
The repetitive stress/release process produces the main periodic
pattern comprising a high peak and a long trough,  apparent
throughout
the gamma-ray light curve.


But we also found that the latter features secondary twin peaks
flanking the main peaks.  The twin peaks are marked by
{lower} and varying amplitudes and by flickers {uncorrelated} with
the main peak's, but also 
{\nnb show} a relatively stable \emph{timing} and a
\emph{smoother} peak/trough pattern.

{\ff The sparing single-jet interpretation in Sect. 4.1 (see Fig.
5, left panel) envisaged twin peaks to be produced again in $j_1$
at the very entering and exiting times of the secondary BH in the
orbital arc of strongest stress. The response of its helicoidal,
magnetized plasma structure induces additional sheared
reconnections and particle acceleration. After onset, the optical,
X-ray and gamma-ray emissions persist for a duration comparable to
the coherent propagation time (a few months)  of the unstable
modes along the jet axis. The plasma column eventually relaxes
back to a lower energy state, with no appreciably enhanced gamma
or X rays. Here the phenomena are turbulent in nature; expected
outcomes include erratic amplitudes and shifting locations of the
twin peaks, yet still flanking a main peak and preserving a
recognizable time distance.}

On the other hand, the two-jet interpretation of Sect. 4.2 took to heart all the above
clues concerning timing, amplitudes and pattern of the twin peaks,
and proposed that a similar stress/release process affects also the
jet $j_2$  carried by the \emph{smaller} binary companion.
However, here the stress takes \ { the specific form of} a
gravitational \textit{torque} exerted by the dominant BH
during the critical orbital arc \ {across the periastron}.
Instability, accelerations and emissions will ensue much as in
$j_1$, but for the reduced amplitudes related to the lower
accretion rates that feed $j_2$.

 Meanwhile, the same torque also
causes the latter to slowly \textit{precede} around an axis nearly
perpendicular to the orbital plane. So $j_{2}$ \ {meets} the
visibility condition (see Fig. 5, right panel) at an axial
inclination $ i < \theta \simeq 1/\Gamma$; this occurs on
{entering } the critical orbital arc, and again on {exiting} it
after about one precessional period. The combined process produces
\textit{twin} minor, closely symmetric peaks flanking the main
peak at a {time} distance around several months, with
\emph{slow} amplitude drift. Occasionally, the nutations {hitherto
averaged out} emerge and drive the emission out of the narrow
visibility range, so that {\nnn the observer misses a twin peaks
event}, as apparently occurred in cycle 3. {\nnn Along this line,}
{\mtt we proposed that } the gravitational force setting the
orbital dynamics also destabilizes the main jet $j_1$ along a
short arc across the periastron, so triggering electron
accelerations with associated S and IC radiations. Meanwhile, on
board the smaller companion $M_2$ the related torque {may affect}
the secondary  {accretion disk} and jet $j_2$. Thus the latter is
induced to emit its own, weaker radiations, and {also to} slowly
precess and direct them in and out the narrow angular range of
visibility.

Overall, we {\nnn suggest} that the binary dynamics  drives the
processes originated in the jets associated with  either binary
component  into 
phase \emph{permanence} and amplitude {shifts}.
{\nnb Single-jet} vs. two-jet interpretations of twin peaks
predict \emph{erratic} changes vs.\emph{smooth} amplitude shifts
from cycle to cycle. In this light, it will be rewarding to keep
under close watch the data evolution from PG 1553+113 over the
next few years.
 In fact, repeated and
shifting twin peaks superposed onto the
 main peak/trough pattern in its
gamma-ray light curves  may well limit the formal confidence {\nnn
in simply periodic behavior} that can be obtained from any
analytical tool.
 On the other hand, evidence of two jets  {\nnn will }
\emph{directly }provide a {double
signature} for the \textit{binary} nature of the SM system
underlying the repetitive emissions of this blazar. {\nnn Such a
signature would bypass the often
contentious issue of 
 red noise contaminating
many searches  for other periodic blazars, see discussion in
\cite{sandrinelli2017}; {\rr see also} \cite{zhang2017} and
\cite{prokhorov2017} who agree with CTV17 on giving PG~1553+113
the standing of a currently primary candidate.}


The importance of finding binary SMBHs in general has been widely
discussed in recent literature since the pioneering work by
\citealt{begelman1980}; to name just a few, we refer  to
\citealt{colpi2014, volonteri2016, klein2016, amaro-seoane2017}.
Here we stress the wide import of establishing even  a single,
relatively nearby and bright binary blazar. This  would also imply
(see Sect. 1) outside the aperture of {\nnn their} e.m. jets many
more {\nnn misaligned and unconspicuous BL Lac candidates at
comparable redshifts,   {\nnn even though only a fraction would be
binaries at} interesting stages of
orbital evolution. Thus even one established
binary blazar will constitute a long stride {toward understanding}
birth and evolution of a class of similar systems, and {setting}
targets to plan searches of low frequency and large amplitude
gravitational waves in the developing project LISA}.

\vspace{0.5 cm}
We acknowledge useful discussions with P. Buratti and F. Pegoraro
on {\nnn tearing} instabilities  {\nnn in columns or tubes of}
laboratory plasmas. {\nnn We thank our referee for constructive
comments that were useful toward a clearer presentation.} Research
partially supported by the ASI grant no. I/028/12/4.













\newpage
\begin{center}
\appendix{}
\end{center}

\section{Fermi-LAT data analysis}


For the purpose of this work, we used the Science Tools provided
by the \textit{Fermi} satellite team\footnote{{\tt
http://fermi.gsfc.nasa.gov}} on the Pass8 data around the position
of \pgp. The version of the Science Tools used was {\tt v10r0p5}
with the standard {\nnn $P8R2\_SOURCE_V6$} instrument response
function (IRF). The reader is referred to  \textit{Fermi}
instruments publications for further details about IRFs and other
calibration details \citep{ackermann2012b}.

We have adopted the current Galactic diffuse emission model ({\tt
gll\_iem\_v06.fits}) in a likelihood analysis and {\tt
iso\_P8R2\_SOURCE\_V6\_v06.txt} as the isotropic model;
the \fermi\ 3rd Point Source Catalog {\tt gll\_psc\_v16.fit} \citep{acero2015} has
 been also used\footnote{{\tt http://fermi.gsfc.nasa.gov/ssc/data/access/}}.
 In modelling the data, the galactic background and diffuse components remained fixed.
 We selected  Pass8 FRONT and BACK transient class events with energies between 0.1 and 300 GeV.
 Among them, we limited the reconstructed zenith angle to be less than 105$^{\circ}$ to greatly
 reduce gamma rays coming from the limb of the Earth's atmosphere. We selected the good time intervals
 of the observations by excluding events that were taken while the instrument rocking angle was larger than 50$^{\circ}$.

For our source PG~1553-113 we adopted a power-law model,
and used the {\tt make3FGLxml.py} tool to obtain a
model for the sources within 25$^{\circ}$ region of inrerest
(ROI). To analyze the data we used the user contributed package
{\it Enrico}\footnote{{\tt https://github.com/gammapy/enrico/}}.

We divided each analysis in two steps: in the first one we leave
free all parameters of all sources within a 10$^{\circ}$ ROI,
while the sources outside this ROI up to 25$^{\circ}$ have their
parameters fixed. Then we run a likelihood analysis using the
Minuit optimizer to determine the spectral-fit parameters and
obtain a fit for all these sources. In the second step, we fix all
the parameters of the sources to the fitted values, except for our
source of interest, and run again the
 likelihood analysis with the Newminuit optimizer to obtain a refined fit. At all times,
 for the central target source PG~1553-113 we kept the spectral normalization free.

We analyzed the data from the whole {\it Fermi}/LAT
mission, starting from 2008-08-04 15:43:36 UTC until 2017-10-04
00:00:00 UTC. For the purposes of this work and comparison
with \citealt{ackermann2015}, we produced a lightcurve by dividing
the whole dataset into 20-day bins. The analysis was carried out
using the assumptions made in this work, and we fixed the photon
index at the value 1.604.

We obtained results compatible with those {\nnn by }
A15, and produced new light curve points beyond the ones {\nnn
given} in that work.

\section{Phase Dispersion Minimization}

To look for periodicity in the {\it Fermi}/LAT data we
used the Phase Dispersion Minimization (PDM) method
\citep{stellingwerf2006}. This computes the variance of the amplitude
on each bin of a data sample. The overall variance of the data set
is compared to the bin variances; if a true period
 is found, the ratio between the bin and the total variances will be small,
 whereas for
 a false period this ratio would be of order 1. This method has
 the advantage of being effective regardless of the shape of the possible period,
 as it is not based onto Fourier transforms of the data. A plot of the amplitude
 versus the trial period will show  the true period {\nnn in the form of a
 deep minimum in the plot}.

To asses the significance of the  period, we used the method by
\citealt{linnell1985}, which is based on Fisher's Method of
Randomization.
 In our case, PDM gives us a best period for our data set of $P = 795.6 \pm 10.0$ days
  ($ 2.18 \pm 0.03$ years).

\section{Continuous Wavelet Transform}

The Continuous Wavelet Transform (CWT) of the Fermi-LAT light
curve reported as a power spectrum in Fig. 3 has been calculated
using the IDL procedure \textit{wavelet.pro} provided by
\citealt{torrence1998} and available at URL:
http://paos.colorado.edu/research/wavelets/. {\nnn In Fig. 3 we
report the map we find in the space: time - period - spectral
power (colors)}.

\end{document}